# Analysis, classification and detection methods of attacks via wireless sensor networks in SCADA systems

[1]Pavel Viktorovich Botvinkin, [1]Valery Anatol'evich Kamaev, [2]Irina Sergeevna Nefedova, [2]Aleksey Germanovich Finogeev, [2]Egor Alekseevich Finogeev

[1]Volgograd State Technical University, Lenina avenue, 28, Volgograd, 400005, Russia
[2]Penza State University, Krasnaya str., 40, Penza, 440026, Russia

**Abstract:** Effectiveness of information security of APCS (automated process control systems), as well as of SCADA (supervisory control and data acquisition) systems, depends on data transmission's protection technologies applied on transport environment's components. This article investigates the problem of detecting attacks on WSN (wireless sensor networks) of SCADA systems. As the result of analytical studies the authors developed the detailed classification of external attacks on sensor networks and brought the detailed description of attacking impacts on components of SCADA systems in accordance with selected directions of attacks. Reviewed the methods of intrusion detection in wireless sensor networks of SCADA systems and functions of WIDS (wireless intrusion detection systems). Noticed the role of anthropogenic factors in internal security threats.
[Botvinkin P. V., Kamaev V.A, Nefedova I.S, Finogeev A.G., Finogeev E.A. **Analysis, classification and detection methods of attacks via wireless sensor networks in SCADA systems.** *Life Sci J* 2014;11(11s):384-388] (ISSN:1097-8135). http://www.lifesciencesite.com. 87

**Keywords:** information security, SCADA system, wireless sensor network, WSN, network attacks, network attacks detection, intrusion detection system, WIDS, anthropogenic security threats.

**Introduction.**

Effectiveness of solving problems of information security of APCS and SCADA systems mainly depends on data transmission's protection technologies applied on transport environment's components.

SCADA systems use a wired or wireless sensor networks as a transport environment for gathering telemetry data and sending commands to executing devices [1, 2].

Because of the transition from wired to wireless networking technologies for the construction of sensor networks for gathering telemetry data, the quality of such protection [3] is determined not only by hardware and software solutions for industrial controllers and sensor nodes, but also by the chosen principles of their information interaction in the process of synthesis of network topology, routing determination and data transfer.

Traditional information security measures (use of sophisticated encryption algorithms, multi-factor authentication, antivirus programs, firewalls, etc.) are not always applicable due to the limited computational and energy resources of sensor nodes and WSN as a whole. In addition, manufacturers of industrial automation and execution devices are developing proprietary protocols, which don't allow implementing security technologies using IPSec, SSL, VPN, etc.

If the SCADA system is set in a large area, for example, for monitoring and management of distributed engineering services (heat, water, electricity and gas supplies) [4, 5], then, as a transport environment often used networks of mobile operators (modem connections GPRS/3G) with the possibility of public access [6, 7]. This effectively provides a channel for attacks.

Therefore, to build an effective ways of protecting information in wireless sensor networks it is necessary to analyze the possible types of attacks, methods of their detection, and reasons of system vulnerabilities. Solving these problems was the purpose of this article.

**Classification of attacks in wireless sensor networks.**

The modern trend towards transport environment of SCADA systems defines the use of self-organizing wireless networks, which feature peer equality, dynamically changing topology, the possibility of network reconfiguration, self-recovery, dynamic routing, etc.

Currently used principles of data transmission in wireless networks provide the possibility of making the four types of impacts: interception, alteration, destruction and code injection. In accordance with the definition of security, all attacks on WSN of SCADA systems can be divided into the following categories:

1) Access attacks, which include attempts to gain unauthorized access to system resources.

2) Attacks on privacy, which represent attempts to intercept the data transfer in the transport environment.

3) Attacks on integrity, which include the generation and transfer of frames to capture and

control of the SCADA system, to call faults and failures in its work or to prepare other attacks.

Consider in detail the classification of attacks by the directions of impacts and give a detailed description of the main types (Fig. 1).

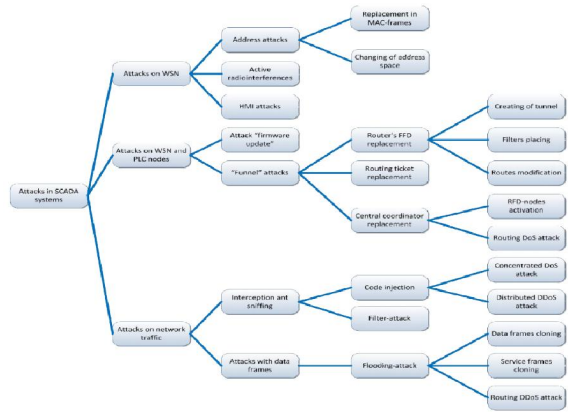

**Fig. 1 Classification of attacks by directions of impacts**

1.　Attacks on sensor network of SCADA system.

1.1.　Creating active interference in the work area of the SCADA system. To create permanent noises "white noise" generators are used. They operate on the same frequency as the SCADA system. A source of that noise can be determined using spectrum analyzers and it's possible to stop the attack by locating and eliminating its source [8]. The most dangerous are natural (lightning) or artificial impulse noises, that can lead not only to a system failure, but also damage the sensor nodes and industrial controllers.

1.2.　Attacks on HMI (human-machine interface) of SCADA system [9]. Unauthorized access to the web-interface from a mobile device can be carried out in the case of open wireless networks or networks with weak authentication.

1.3.　Attacks on WSN addresses changing (spoofing), aimed on initiating DoS (Denial of Service). We can distinguish two types of such attacks:

1.3.1.　Interception of sensor nodes' frames with the purpose of spoofing MAC source and destination addresses, which leads to the failure or malfunction of the SCADA system.

1.3.2.　Replacement of central coordinator to change the address space in configuration of sensor network.

2.　Attacks on sensor network's nodes and related devices.

2.1.　Changing the firmware, drivers and software of industrial controllers (PLC — Programmable Logic Controller) and terminal sensor nodes (RFD — Reduced Function Device). Attack conducted by scanning PLCs and RFDs to identify opportunities of changing preinstalled operating system, firmware, drivers and controllers.

2.2.　Injection attacks by spoofing or replacement the nodes of WSN, responsible for collecting and relaying data in a network (FFD — Fully Function Device) to intercept and redirect network traffic. The main purpose of such an attack is to redirect network traffic to an injected or replaced node. Consider the variety of such an attack:

2.2.1.　Compromising the node by replacing routes confirmation tickets to redirect traffic from the end source-nodes to an injected receiver-node [10]. As a result of such replacement, the real coordinator stops collecting data from the PLCs and sensors, and dispatch service loses control of technological processes.

2.2.2.　Router (FFD node) replacement in a sensor network aimed to violate the correct operation of routing algorithms. Attack can be carried out by:

—　creation of a "false" tunnel (on the injected router runs a program that copies retransmitted frames to transfer them to another sensor network, or, conversely, the program, the frame control commands from another network);

—　setting filters (on the injected router runs a program that filters and destructs retransmitted frames on the specified criteria or content);

—　changing routes (on the injected router runs a program that changes the contents of "Route Record" packets by a given algorithm or randomly.

1.1.1.　Replacement of a central coordinator of organization's WSN aimed to run-up broadcast storm and to achieve denial of service or to fast discharge the power supplies [11].

2.　Attacks on sensor network's traffic.

2.1.　Listening of data transmission channels. Produced by intercepting and decoding network traffic with special utilities (sniffers) for subsequent analysis of frames for extracting the required information.

2.2.　Attacks with data frames. Performed by flooding or by generating "false" service or data frames or replace the contents of captured frames and the subsequent injection into the network. Consider the basic options such attacks.

2.2.1.　Injection of malicious code. After the injection of malicious code, it focuses on bringing malfunction to the executing devices, the entire network or on changing the parameters of technological processes. Injection into the network routers a self-replicating worm leads to infection and transformation of all nodes to bot-net, whose nodes generate data frames to increase the reaction time of the network, producing faults and failures (distributed DoS attack [12, 13]).

2.2.2. Frames filtering and selective forwarding. Produced by injecting into network a special type of software or hardware filters that intercept data frames, filter them, and may perform a custom broadcasts. Effectiveness of the attack increases with its integration with the "funnel" attack.

2.2.3. Flooding attacks by generating "false" service or data and broadcasts:

— cloning and broadcast the data frames (performed by intercepting and repeatedly reproducing the same data frames followed by broadcasting on the network to achieve input buffers overflow and network failures;

— generation and broadcast frames of polling units and HELLO-frames to achieve failures of network resources (creating and sending in the network set HELLO-frames with non-existent addresses of nodes, it's possible to make an image of "non-existent" area of the sensor network;

— synthesis of "virtual" source-nodes to broadcast route packets from them (routing DdoS attack, here is exploited the weakness of Source-Routing technology if it's used in centralized SCADA systems with one coordinator and gateway — excessive network load with broadcast routing traffic).

**Intrusion detection in wireless sensor networks of SCADA systems.**

There are few general traditional techniques aimed to detect attacks on transport network media [14, 15]. All of them include the following procedures:

— identification and validation of non-standard network traffic;

— periodic inspection of privileges and authorizations for personnel access to specific information resources of SCADA system;

— disabling the unused protocols and services;

— disabling the remote access and control to the network nodes and applications;

— periodically scans of network interfaces and drivers;

— timely software updates for nodes from trusted source.

There are three ways to detect attacks in networks:

1. Detection by the signatures. The signature defines the characteristics (profiles) of previously committed attacks. During the scanning reveals a coincidence of signatures and notification is made. However, this method does not reveal the attack with new (unknown) signatures.

2. Detection of the anomalous behavior. Attack detection occurs at identifying abnormal behavior of the network node or deviations from its normal operation. The disadvantage of this approach is that incorrect behaving node may be affected by other factors that are not related to the attacks, such as software, hardware or sensor failure.

3. Combined detection by the specifications. This method combines the two previous ones to reduce their shortcomings.

WIDS (Wireless Intrusion Detection System) is a software and hardware solution, which consists of software agents that perform the function of collecting, processing and analyzing network traffic packets. Agents interact with the server, transmitting captured packets to it. Server processes the received data to detect attack signatures and anomalous behavior of network nodes, as well as responding to events. Thus, WIDS combines signature and behavioral ways, and relates to the third method. In operation, WIDS performs monitoring and analysis of traffic in sensor network. Its functionality includes the following standard procedures.

1. Analysis of the topology of WSN.

2. Determination of WSN's vulnerabilities.

3. Compilation and maintenance of lists of network nodes. These lists are generated on the basis of network traffic analysis and retrieval of MAC-addresses of the network nodes from captured frames. The resulting lists in the future actually allow detecting the appearance in the network new "foreign" potentially dangerous nodes.

4. Detecting and countering attacks in WSN. At the moment, the number of detected attacks in WSN is far less than the number of detected attacks in wired networks, as it is limited only by the analysis of traffic data on the data link layer of the OSI model. The result of the detection of the attack is the administrator's notification of potential problems in different ways (via email, SMS messages, etc.) and event logging for auditing.

5. Locating the source of the attack and its suppression. WIDS can use such mechanisms of repression, as the implementation of DoS attacks on attacker's node, blocking the attacking agent by active network equipment. Locating the source of attacks means detection of the coordinates of the device that violates security policy by the trilateration, multilateration or triangulation technologies.

6. Control of security policy. Based on the analysis of the network nodes list in order to detect changes in the policies set by the administrator. An audit can detect the appearance of unauthorized nodes and applications, violations of traffic protection policy.

7. Performing controlled invasion tests through existing vulnerabilities of SCADA system and its components by specific exploits.

8. Monitoring wireless network capacity and network response time. In the process of monitoring, WIDS can monitor the physical and data link layers of the network, and identify problems such as:

— overload of channel, node or network,

— a sharp increase in the number of data frames received by the coordinator, routers and end nodes,

— reduction of power of radio signals,

— sharp increase in the broadcast service or routing frames,

— overlap with neighbor networks,

— reduction in network bandwidth for no apparent reason,

— dramatic increase of route search time,

— sharp increase of server applications' reaction time to client requests,

— increase of collisions in data channels,

— appearance of new network nodes,

— reduction of the data transmission rate,

— overload the network nodes and the network as a whole,

— overflow of buffer memory nodes, denial of service, etc.

On the basis of analysis of the results of such monitoring by persons responsible for the information security, the necessary decisions may be concluded and appropriate operational and long-term measures may be implemented.

**Conclusion.**

Despite the fairly large number of possible attacks on wireless sensor networks and SCADA systems, the most dangerous are the internal anthropogenic threats to information security, which include:

— unintentional personnel actions that create the auspicious conditions for external attacks by hackers,

— intentional ignoring the requirements of information security by staff that serving SCADA system,

— the lack of qualification of personnel in the field of information technologies and implementation of methods of information security.

Unlike external intruder, the staff of the enterprise has the great opportunities for attacks to infect and spread malicious code on the sensor network. Information security problems often caused not so much by external attacks, but more as non-compliance with staff regulations and rules of the enterprise information security policy. Managers and other staff of the enterprise may ignore their duties and in the "free" time do the Internet "surfing", social networking, and playing computer games. The result may be an unauthorized PC infection by computer viruses, Trojan horses and worms, which then may penetrate into the sensor networks. This explains the fact that viruses and worms like Stuxnet often present in industrial systems, and the fact that their presence is normally hidden by staff and managers, as the disclosure of this information will lead all the staff and management to the detailed inspection and then to subsequent negative consequences for them. In addition, the finding of the infection in the SCADA system may cause a need of hard reset to clean the virus and will stop the most of enterprise's processes, but it is not always feasible from an economic standpoint.

Also, the lack of qualifications of personnel which works with PLCs and SCADA systems requires the involvement of outside experts to identify and correct software changes in controllers, because after cleaning the system it's necessary to be ensured that the programs and settings in the controllers [16] correspond to the values required for the proper functioning of complex of industrial automation.

It is well known that the human factor is the main reason of deviations from normal operation status in various technical systems. This requires special attention to the establishment and maintenance of appropriate technical regulations.

This article was financially supported by the Russian Ministry of Education in scope of the base part (project 2586 of task #2014/16).

**Corresponding Author:**
Dr. Botvinkin Pavel Viktorovich,
Volgograd State Technical University, Lenina avenue, 28, Volgograd, 400005, Russia

7/9/2014